\documentclass{amsart}

\usepackage{amsmath}
\usepackage{amssymb}
\usepackage{amsthm}
\usepackage{graphicx}
\usepackage[table]{xcolor}
\usepackage{multirow}

\newcommand{\diff}{\mathrm{d}}
\newcommand{\R}{\mathbb{R}}
\newcommand{\N}{\mathbb{N}}
\newcommand{\bE}{\mathbb{E}}
\newcommand{\cD}{\mathcal{D}}

\newcommand{\etaName}{stability index}

\newtheorem{theorem}{Theorem}
\newtheorem{examp}[theorem]{Example}

\definecolor{backgroundboxcolor}{gray}{0.85}

\newenvironment{backgroundbox}{
\vspace{0.2cm}
\noindent
\begin{tabular}{p{0.97\textwidth}}
\rowcolor{backgroundboxcolor} 
\vspace{-0.1cm}
}{
\vspace{0.1cm}
\end{tabular}
\vspace{0.2cm}
}

\begin{document}

\title[Stochastic Reaction-Diffusion Systems in Biophysics]{Stochastic Reaction-Diffusion Systems in Biophysics: Towards a Toolbox for Quantitative Model Evaluation}
\author[G. Pasemann, C. Beta, W. Stannat]{Gregor Pasemann\footnotesize{$^{1}$}, Carsten Beta\footnotesize{$^{2}$} and Wilhelm Stannat\footnotesize{$^{3}$}}
\email{gregor.pasemann@hu-berlin.de \\ beta@uni-potsdam.de \\ stannat@math.tu-berlin.de}
\date{\today}

\maketitle

\footnotetext[1]{Institut f\"ur Mathematik, Humboldt Universit\"at zu Berlin, Unter den Linden 6, 
10099 Berlin}
\footnotetext[2]{Institut f\"ur Physik und Astronomie, Universit\"at Potsdam, Karl-Liebknecht-Stra{\ss}e 24/25, 14476 Potsdam}
\footnotetext[3]{Institut f\"ur Mathematik, TU Berlin, Stra{\ss}e des 17. Juni 136, 10623 Berlin}

\begin{abstract}
We develop a statistical toolbox for a quantitative model evaluation of 
stochastic reaction-diffusion systems modeling space-time evolution of 
biophysical quantities on the intracellular level. Starting from 
space-time data $X_N(t,x)$, as, e.g., provided in fluorescence 
microscopy recordings, we discuss basic modelling principles for 
conditional mean trend and fluctuations in the class of stochastic reaction-diffusion systems, and subsequently develop statistical inference methods 
for parameter estimation. With a view towards application to real data, we 
discuss estimation errors and confidence intervals, in particular in 
dependence of spatial resolution of measurements, and investigate the 
impact of 
misspecified reaction terms and noise coefficients. We also briefly touch 
implementation issues of the statistical estimators. 
As a proof of concept we apply our toolbox to the statistical inference on 
intracellular actin concentration in the social amoeba \textit{Dictyostelium 
discoideum}. 
\end{abstract}

\section{Introduction}
\label{sec:1}

\subsection{A Motivating Example}

\begin{figure}[t]
    \begin{center}
    \includegraphics[height=0.3\textheight]{"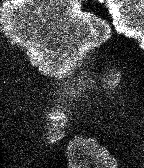"}
    \includegraphics[height=0.3\textheight]{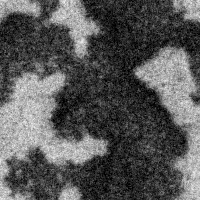}
    \end{center}
    
    \caption{
    Actin marker concentration in \textit{D. discoideum} giant cells: Experimental data (left) and simulation using a FitzHugh--Nagumo type model proposed by Sergio Alonso \cite{FlemmingFontAlonsoBeta20} (right). 
    {Experimental data provided by Sven Flemming.}
    }\label{fig:example}
\end{figure}

Reaction-diffusion systems are one of the most popular classes of models in the life sciences~\cite{murray_mathematical_2003,edelstein-keshet_mathematical_2005}.
They arise as a natural description of spatially extended systems, where the dynamics at each location can be represented by a reaction kinetics that is often nonlinear, while neighboring locations in the system are coupled by diffusive transport of one or several of the reacting species.
Such models are used to approximate the space-time dynamics of diverse systems and on vastly different length scales, including models of vegetation patterns~\cite{meron_nonlinear_2015}, animal skin and tissue~\cite{kondo_reaction-diffusion_2010}, as well as subcellular dynamical structures~\cite{beta_intracellular_2017}.
As most of these systems are inherently noisy, such models are often extended by including noise terms into the reaction kinetics, thus turning them into stochastic reaction-diffusion systems.
Here, we will focus, as a motivating example, on intracellular wave patterns that emerge in many different cell types and have been associated with essential cellular functions, such as motility, phagocytosis, and cell division~\cite{inagaki_actin_2017}.
To date, it is not clear, whether the same generic mechanism drives the formation of waves in different cell types, or whether fundamentally different model structures are required to address the observations across different species.
Also on the molecular level, different players were found to be essential for wave formation in different cell type.
Nevertheless, in most cases, the protein actin, which serves as the building block of one of the major classes of cytoskeletal filaments and is highly conserved across all eukaryotic cell types, is associated with wave formation, so that these waves are commonly denoted as ``actin waves''~\cite{allard_traveling_2013,beta_actin_nodate}.

A widely used model organism to study the dynamics of actin waves is the social amoeba {\it Dictyostelium discoideum} ({\it D.~discoideum})~\cite{annesley_dictyostelium_2009}.
Since the first observations of actin waves in {\it D.~discoideum} in the 1990s, their dynamics and molecular composition have been studied in great detail~\cite{vicker_pseudopodium_1997,gerisch_mobile_2004,arai_self-organization_2010}.
In the wake of this experimental progress, many models have been proposed to describe the dynamics of actin waves in {\it D.~discoideum}.
These models are typically of reaction-diffusion type, with some of them also including noise terms.
However, their level of complexity may vary widely, mostly with respect to the number of signaling components included in the reaction terms.
The more biologically oriented models range from 4 or 5 components up to dozens of signaling molecules. 
In contrast, also minimal models have been proposed with the aim of capturing the essential dynamical features of the intracellular wave phenomena with the most minimal set of mathematical equations.
These models are often inspired by generic pattern forming reaction-diffusion systems, such as the FitzHugh-Nagumo, Gray-Scott, or Brusselator models~\cite{beta_actin_nodate}.

While numerical simulations of many of these models show close agreement with experimental recordings of actin waves, it remains an open question, which of the models performs best in the light of available data.
How robust is the agreement with experimental data?
Are the model parameters realistic? 
Which are the essential nonlinearities and coupling terms that need to be included in the model?
Developing mathematically rigorous approaches to infer parameter values as well as the structure of reaction terms and noise contributions in stochastic reaction-diffusion models from experimental data will greatly contribute to a more systematic model design to advance our understanding of intracellular actin waves and their associated biological functions.

As a prototypical example of quantitative model evaluation in a stochastic reaction-diffusion system modelling the space-time evolution of intracellular waves, we will consider here
a spatially extended stochastic FitzHugh-Nagumo-type model of actin wave formation in {\it D.~discoideum}~\cite{PasemannFlemmingAlonsoBetaStannat21}.
The model was initially introduced to describe wave-induced cytofission events in {\it D.~discoideum}~\cite{FlemmingFontAlonsoBeta20} and was developed by extending a model of amoeboid motility based on a noisy bistable switch~\cite{alonso_modeling_2018,moreno_modeling_2020}.
In order to assess the quality of this model, beyond a visual comparison of simulation output with measurements, a crucial step is to estimate the unknown model parameters together with corresponding confidence intervals. 
To this end, a mathematical theory for parameter estimation of stochastic reaction-diffusion systems has been developed in~\cite{PasemannFlemmingAlonsoBetaStannat21} and was applied primarily to the estimation of the effective diffusivity of the actin concentration/density and the estimation of other reaction constants.  

The data in this example was given in terms of confocal laser scanning microscopy recordings of giant {\it D.~discoideum} cells generated by electric pulse-induced fusion of individual cells~\cite{gerisch_membrane_2013}.
The use of giant cells provides a setting, where wave dynamics can be recorded over extended cortical areas without having to take boundary effects into account and where system size effects can be systematically explored~\cite{gerhardt_actin_2014,yochelis_versatile_2022}.
Quantitative parameter estimates like this provide first opportunities to perform sanity checks on a proposed model by comparing the obtained numbers with know literature values or physical limits in their order of magnitude.
We will turn to the results of the statistical analysis of this example in more detail in Section~\ref{sec:Application} below.

\subsection{Context and Outline}

Most of the existing literature on the application of reaction-diffusion 
systems deals with modelling, numerical approximation and (statistical) 
analysis of simulation output \cite{Bressloff23, Lord14}. 
In fact, 
identifying and validating differential equations explaining the space-time 
evolution of physical observables is a classical problem in the natural and 
the engineering sciences and there is a huge body of literature devoted to 
data-driven approaches, see in particular 
\cite{Sindy2016} and the references therein. 
When it comes to the explanation of 
space-time data obtained in real experiments however, there is little 
existing literature on how to perform a quantitative validation of these 
models. This is partly due to the fact that the statistical 
inference of stochastic reaction-diffusion systems, and more general 
stochastic partial differential equations (SPDEs), is far from being well-developed. 
Another reason is that space-time data of sufficiently fine space time 
resolution had not been available or not so easily tractable in the past. 
Notably, there are early examples \cite{Unny89, KumarUnnyPonnambalam91} of SPDE-based data analysis in the field of groundwater hydrology by means of methods that are very close to ours. More recently, a cell repolarization model was considered in \cite{AltmeyerBretschneiderJanakReiss22}. 
We believe that a theoretical foundation of the statistical methods in use can improve the understanding of phenomena observed in the data, while the interplay between experiments, simulations and theoretical analysis may also lead to fruitful new directions within mathematical research. 
\\

It is the main purpose of this work to put the approach persued in \cite{PasemannFlemmingAlonsoBetaStannat21} into the larger perspective of developing statistical tools for the quantitative model evaluation of stochastic-reaction diffusion systems in biophysics. 
We 
structure this work as follows: In Section \ref{sec:DynamicalModel} we 
discuss how the dynamics of a random field $X(t, x)$ is modelled as a 
stochastic partial differential equation, and discuss different commonly used terms for the drift and the noise. In Section \ref{sec:StatisticalModel}, we 
provide a method to estimate unknown parameters in such a model. Section 
\ref{sec:StatProp} is devoted to structural properties of the resulting 
estimators that can be proven with mathematical rigour. In particular, we 
discuss the principal differences between diffusion and reaction terms, the 
high resolution limit vs. the large time limit, and the impact of model 
misspecification. In Section \ref{sec:Implementation}, we comment on the 
time discretization of some integrals appearing in the estimation method. 
Finally, Section \ref{sec:Application} shows an application of our method to 
\textit{D. discoideum} giant cell data.

\section{The Dynamical Model --- Stochastic Reaction--Diffusion Equations}\label{sec:DynamicalModel}

\subsection{General Considerations for Space-Time Random Fields}

We start with discussing general principles for the derivation and 
further assessment of stochastic evolution equations
\begin{equation}
    \label{Eveq1}
    \partial_t X = \mathcal{F} (X) + \xi 
\end{equation}
as mathematical models describing space-time evolution of a given 
physical observable. Here, $X$ is thought of as a random 
field in space time $X(t,x)$, and the basic decomposition 
\begin{equation} 
\label{InfMean} 
\mathcal{F} (X)(t,x) \approx \frac 1h \bE\left[ X(t + h, x) - X(t,x) 
\mid X(s, \cdot ) , s\le t  \right] , 
\end{equation} 
describing the mean trend in the time evolution of $X$, given the 
information of the past, $X(s, \cdot)$, $s\le t$, and 
\begin{equation} 
\label{noise} 
\xi (t,x) \approx \frac 1h \left( X(t+h,x) - X(t,x) -\mathcal{F} (X)
(t,x) h\right) 
\end{equation} 
models the corresponding remainder as fluctuations in terms of a 
stochastic process $\xi (t,x)$. Besides theoretical 
considerations on the overall structure of the drift term  
$\mathcal{F}$ and the noise term $\xi$ on the r.h.s. of \eqref{Eveq1}, 
our aim here is to provide additional statistical tools for a quantitative 
model evaluation of \eqref{Eveq1}, given space time data $X_N(t, x)$ 
whose time-evolution should be explained by \eqref{Eveq1}.

In principle, a first expansion of $\mathcal{F} (X)$ can 
be achieved separating linear from nonlinear components. A simplifying 
assumption is that the linear part dominates the nonlinear part. What 
''dominating'' means will be made precise in Subsection  
\ref{sec:robust:reaction}. In this work, we restrict to the widely used 
case $\mathcal{F}(X) = \vartheta_0\Delta X + F_\vartheta(X)$, i.e. spatial 
interaction within the drift is governed by an isotropic diffusion term 
of intensity $\vartheta_0$, and $F_\vartheta(X)$ contains lower order (e.g. 
reaction) terms present in the drift, as explained in the next section.
A common approach to obtain the noise term $\xi (t,x)$ which models the 
(conditional) fluctuations is to match the covariance of 
$\frac 1h \left( X(t+h,x) - X(t,x) -\mathcal{F} (X) (t,x) h\right)$ 
to that of a Gaussian process (additive noise), or to additionally weight such a process with $X$-dependent coefficients (multiplicative noise).

\subsection{Stochastic Reaction-Diffusion Models}

Based on these considerations, let us next fix the precise mathematical setting for our analysis. Our 
quantity of interest is a random field $X(t,x)$, 
where $t\in[0,T]$ and $x\in\cD$ for some domain $\cD\subset\R^d$.  
We restrict to the case that $\cD=[0,L_1]\times\dots\times[0,L_d]$ is a $d$-dimensional 
rectangle with periodic boundary conditions. 
We assume that the temporal evolution of $X$ is governed by three components:
\begin{align}\label{eq:model}
    \partial_t X = \vartheta_0\Delta X + F_\vartheta(X) + \xi
\end{align}
\begin{itemize}
    \item $\vartheta_0\Delta X$ describes the diffusive coupling, $\Delta=\sum_{i=1}^d\partial_{x_i}^2$ being the Laplacian operator, and $\vartheta_0>0$ the diffusivity of the process. We highlight this term for two reasons: First, it seems to be a common feature of most (S)PDE models that are used in practice. On the other hand, it plays a distinguished role when it comes to statistical inference, as explained in Section \ref{sec:StatProp}.
    \item $F_\vartheta(X)$ includes all other explicitly modelled dynamical forces acting on the temporal evolution of $X$. 
    Examples are given below. 
    $\vartheta\in\R^p$ is a vector of parameters on which $F$ may depend (e.g. reaction rates).
    \item $\xi=\xi(t, x)$ denotes random fluctuations, which may be either an inherent property of the dynamics, or may model an approximation of unresolved fast-scale components. We do not impose assumptions on $\xi$ apart from statistical ones: In the simplest case, $\xi$ is just space-time white noise, but additional assumptions on the spatial or temporal correlation structure can be made.
\end{itemize}

\subsection{Models for the Drift}

\begin{examp}[reaction-advection-diffusion equation]
    Consider \label{ex:reacadvdiff}
    \begin{align}
        \partial_tX = \vartheta_0\Delta X - \nabla\cdot(Xv) + f(X) + \xi,
    \end{align}
    where $v(t, x)$ is a transport velocity field and $f:\R\rightarrow\R$ a function describing local sources and sinks. Write $F_\mathrm{adv}(X)=-\nabla\cdot(Xv)$ for the advection term and $F_\mathrm{reac}(X)=f(X)$ for the source term, then this is a special case of \eqref{eq:model} with $F=F_\mathrm{adv}+F_\mathrm{reac}$.
\end{examp}

\begin{examp}[activator-inhibitor system]
    Consider the 
    system of FitzHugh--Nagumo type \cite{FlemmingFontAlonsoBeta20, PasemannFlemmingAlonsoBetaStannat21} \label{ex:actinh}
    \begin{align}
        \partial_t U &= D_U\Delta U + k_1 U(u_0-U)(U-u_0a) - k_2 V + \xi, \label{eq:FN:act} \\
        \partial_t V &= D_V\Delta V + \varepsilon(bU-V), \label{eq:FN:inh}
    \end{align}
    with $U(0, x)=V(0, x)=0$ for simplicity.
    Here, $U$ is the activator, and $V$ is the inhibitor. Later, we will treat the latter as an unobserved latent component. This model can be used to generate various dynamical patterns, including traveling waves, bistability and excitability. $D_U,D_V>0$ are the diffusivity constants of $U$ and $V$, whereas $k_1,k_2, \varepsilon, b>0$ denote reaction rates, and $u_0>0$ and $0<a<1$ determine the stable and unstable fixed points of the reaction model. This system can be cast into the form \eqref{eq:model}: Note that by the variation of constants formula, $V$ admits an explicit representation
    \begin{align}\label{eq:inhRep}
        V(t, x) = \varepsilon b\int_0^t\int_\mathcal{D}e^{-\varepsilon(t-s)}G(D_V(t-s), x-y)U(s, y)\diff y\diff s,
    \end{align}
    where $G(t, x)$ is the Green's function for the heat equation on $\mathcal{D}$.
    In particular, $V$ is determined by the activator history up to time $t$: We write $V=F_{\mathrm{inh}}(U)$. Additionally, let $F_\mathrm{act}(U)=f_\mathrm{act}(U)=k_1U(u_0-U)(U-u_0a)$. Then the activator component $U$ satisfies an equation of the form \eqref{eq:model} with $F=F_\mathrm{act}+F_\mathrm{inh}$.
\end{examp}

\begin{examp}[spatiotemporal coupling term]
    Extending the previous example, we can consider moving average-type interactions of the form
    \begin{align}\label{eq:Fmovavg}
        F(X)(t, x) = \int_0^t\int_\cD K(t-s, x-y)X(s, y)\diff y\diff s,
    \end{align}
    where $K(t, x)$ is some kernel function. If $K(t, x)=e^{-\varepsilon t}G(D_Vt, x)$ with the Green's function $G$ for the heat equation on $\cD$, then we recover the integral term in \eqref{eq:inhRep}.
\end{examp}

\begin{examp}[further drift models]
    There is a large number of frequently used dynamical models that we did not discuss so far. If the velocity field of an advection term is further specified to be of the form $v(t, x) = -\int_\cD\nabla W(x-y)X(t,y)\diff y$ with some interaction potential $W$, the resulting so-called aggregation term can describe non-local attraction and repulsion forces caused by regions of high concentration of $X$.
    Diffusion-aggregation models arise as continuum limits of interacting particle systems, describing time evolution of systems at microscopic levels, and may exhibit intricate dynamical features such as blow-up in finite time, see \cite{CarrilloCraigYao19} for a discussion. 
    Non-Markovian temporal interaction may also be modelled by delay terms.
\end{examp}

\subsection{Models for the Fluctuation}

Finally, we discuss different noise models:

\begin{examp}[space-time white noise]
    If there is no spatial or temporal correlation in the noise, the law of $\xi$ is determined by 
    $\bE[\xi(t, x)\xi(s, y)] = \sigma^2\delta(t-s)\delta(x-y)$, where $\sigma>0$ denotes the noise intensity.
\end{examp}

\begin{examp}[Ornstein--Uhlenbeck noise, first approach]
    Here, temporal correlation is introduced by modelling $\xi$ as a stochastic process in time: \label{ex:noise:OUone}
    \begin{align}\label{eq:OUnoise}
        \partial_t\xi = -\mu\xi + \Xi
    \end{align}
    without loss of generality with $\xi(0, x)=0$, where $\Xi(t, x)$ is a space-time white noise, and $\mu>0$. In this model, $\xi$ is white in space and has an exponentially decaying 
    correlation in time.
\end{examp}

\begin{examp}[Ornstein--Uhlenbeck noise, second approach] Alternatively, temporal correlation may be introduced by letting $W(t, x)=\int_0^t\xi(s, x)\diff s$ instead of $\xi(t, x)$ satisfy the Ornstein--Uhlenbeck equation \eqref{eq:OUnoise}. Compared to Example \ref{ex:noise:OUone}, the solution field $X$ then exhibits different qualitative features. For example, $X$ will not be differentiable in time (it is ``rougher''). \label{ex:noise:OUtwo}
\end{examp}

\begin{examp}[Spatially correlated noise]
    We can include spatial correlation by taking spatial averages:
    \begin{align}
        \xi(t, x) = \int_\cD b(x,y)\Xi(t, y)\diff y
    \end{align}
    with a space-time white noise $\Xi$ and a kernel $b$.
    A typical choice is given by
    \begin{align}\label{eq:kernelFracLaplacian}
        b_{-\gamma}(x, y) = \sum_{k=1}^\infty\lambda_k^{-\gamma}\Phi_k(x)\Phi_k(y),
    \end{align}
    where the $\Phi_k$ are the eigenfunctions of the Laplacian with eigenvalue $\lambda_k$ on the (bounded) domain $\cD$, and $\gamma>0$ is a tuning parameter that governs the correlation strength. Furthermore, $\gamma$ is related to the smoothness of the solution field $X(t, x)$ for fixed $t$, with larger values of $\gamma$ leading to smoother solutions.
\end{examp}

\section{Statistical Model --- Inference on Model Parameters}\label{sec:StatisticalModel}

\subsection{Parametrization of the Reaction Term}

Our standing assumption is that $F$ is known up to a finite-dimensional parameter vector $\vartheta$. Consider the case that $F$ depends linearly on its unknown parameters:

\begin{align}\label{eq:Flinearmodel}
    F_\vartheta(X) = \vartheta_1F_1(X) + \vartheta_2F_2(X)+\dots+\vartheta_pF_p(X) + F_*(X),
\end{align}
where the $F_i$ form a dictionary of components of the reaction term with unknown intensity, 
and $F_*$ represents any ``fixed'' term with known intensity (it may be zero). 
The dictionary terms may arise from some truncated basis expansion, or encode prior knowledge on the dynamics (such as terms generating bistable behaviour, or even possible non-local or non-Markovian terms). 

\begin{backgroundbox}
    We highlight that although the dynamics of $X$ may be highly \textit{nonlinear}, as explained in the previous section, assumption \eqref{eq:Flinearmodel} will lead to a \textit{linear} statistical model. 
\end{backgroundbox}

\begin{examp}[reaction-advection-diffusion equation -- continued]
    In the setting of Example \ref{ex:reacadvdiff}, the velocity field $v$ and the reaction function $f$ may be expanded as (or approximated with) a linear combination of known basis functions with unknown coefficients, e.g. $f(X)=\sum_{k=1}^p\vartheta_kf_k(X)$. If the $f_k$ have a physical meaning, the $\vartheta_k$ have a natural interpretation of reaction rates of different forcing terms.
\end{examp}

\begin{examp}[activator-inhibitor system -- continued]
    In the setting of Example \ref{ex:actinh}, one may argue that the level $u_0$ of the stable phase should be treated as known. In this case, the general scheme \eqref{eq:Flinearmodel} reduces to
    \begin{align}\label{eq:actinh:parametrization}
        \begin{matrix}
            \vartheta_1 = k_1u_0a, & F_1(U) = -U(u_0-U), \\
            \vartheta_2 = k_1, & F_2(U) = U^2(u_0-U), \\
            \vartheta_3 = k_2\varepsilon b, & \quad\quad F_3(U)(t, x) = -\int_0^te^{(t-s)(D_V\Delta - \varepsilon\mathrm{Id})}U(s, x)\diff s.
        \end{matrix}
    \end{align}
    On the other hand, if $u_0$ is treated as unknown, we can still expand $F_\mathrm{act}$ in powers of $U$, i.e. we write $F_\mathrm{act}(U)=\vartheta_1U+\vartheta_2U^2+\vartheta_3U^3$ (where the $\vartheta_k$ are different as in \eqref{eq:actinh:parametrization}) and estimate these coefficients together with the intensity of $F_\mathrm{inh}(U)$.
    However, treating the inhibitor parameters $\varepsilon, D_V$ as unknown, we would leave the class of linearly parametrized statistical models determined by \eqref{eq:Flinearmodel}. An alternative approach is to consider the moving average representation \eqref{eq:Fmovavg} instead and approximate the kernel $K$ therein by a linear combination of basis functions with unknown coefficients.
\end{examp}

\subsection{Observing an Approximation to $X$}

We want to infer the unknown parameters (including the diffusivity) $\vartheta_0,\vartheta_1,\dots,\vartheta_p$ from one realization of $X(t, x)$ for $0\leq t\leq T$ and $x\in\cD$. In fact, we will assume that we have only some approximation $X_N(t, x)$ of $X(t, x)$ at hand. 
This is a more realistic assumption than having all of $X$ available.
There are different ways to model the relation between $X_N$ and $X$. Common approaches are:
\begin{itemize}
    \item \textbf{Spectral approach:} The first $N$ eigenmodes of $X$ are observed, i.e. 
    \begin{align}\label{eq:spectralapproach}
        X_N(t, x)=\sum_{k=1}^N\left(\int_\cD X(t, y)\Phi_k(y)\diff y\right)\Phi_k(x),
    \end{align}
    where the $\Phi_k$ are the eigenfunctions of the Laplacian $\Delta$ on $\cD$. \cite{HubnerKhasminskiiRozovskii93, HuebnerRozovskii95}
    \item \textbf{Local approach:} One (or more) local averages of $X$ are observed, i.e. $X$ is tested against a point-spread function located at $x_0\in\cD$: In this case 
    \begin{align}
        X_N(t)=\int_\cD X(t, y)h^{-1}K(h^{-1}(y-x_0))\diff y,
    \end{align}
    where $K(x)$ denotes the point-spread function and $h=1/N$ is a bandwidth parameter, modelling high precision as $N\rightarrow\infty$. \cite{AltmeyerReiss21} Note that in the case of a single local observation concentrated at $x_0\in\cD$, the process $X_N(t, x)=X_N(t)$ does not even depend on $x\in\cD$. 
    Multiple local measurements can in principle be combined with some interpolation method to fit into the present framework (see also \cite{AltmeyerTiepnerWahl22}).
    \item \textbf{Discrete approach:} $X_N$ is the interpolation of evaluations of $X$ on some discrete grid in space.
    An estimator of least squares type as derived below has been studied in \cite[Chapter 4]{Pasemann21}. For different approaches for discretely sampled data see e.g. \cite{PospisilTribe07, CialencoHuang20, Chong20, HildebrandtTrabs21, HildebrandtTrabs21_nonparametric, KainoUchida21}.
\end{itemize}

Apart from $X_N$ alone, in order to exploit the dynamics of \eqref{eq:model} we also need suitable approximations of the drift terms appearing on the right-hand side of that equation,\footnote{In fact, for constructing the estimator, we only need the terms appearing in \eqref{eq:modelApprox} instead of the approximate process $X_N$ itself.} so that we can write the dynamics of $X_N$ (up to a possible approximation error) as

\begin{align}\label{eq:modelApprox}
    \partial_tX_N = \vartheta_0(\Delta X)_N + \vartheta_1F_1(X)_N + \dots + \vartheta_pF_p(X)_N + F_*(X)_N + \xi_N.
\end{align}

For example, for any reaction term given by a function $F_i=f_i:\R\rightarrow\R$, it may be reasonable to set $F_i(X)_N=F_i(X_N)$, i.e. to evaluate that function on the approximate process. However, terms involving differential operators should be handled with more care, because one needs to approximate derivatives here, which may lead to numerical challenges. In particular, this applies to the diffusion term.\footnote{We note that within the spectral approach, it holds $(\Delta X)_N=\Delta (X_N)$ by design, such that this problem is avoided there.}

\subsection{The Least Squares Method}

Under the assumption of linear parametrization, the problem of estimating the unknown parameters in \eqref{eq:modelApprox} has apparent structural similarities to classical linear regression, $\partial_tX_N-F_*(X)_N$ being the response variable, and $(\Delta X)_N,F_1(X)_N,\dots,F_p(X)_N$ being the predictors. If the dynamic noise $\xi$ is space-time white noise, this suggests a formal least squares method

\begin{align}\label{eq:leastsquares}
    \int_0^T\int_\cD\left(\partial_tX_N - \vartheta_0(\Delta X)_N-\sum_{i=1}^p\vartheta_iF_i(X)_N-F_*(X)_N\right)^2\diff x\diff t\quad\rightarrow\quad\mathrm{min},
\end{align}

and the resulting estimator $\hat\vartheta_N=(\hat\vartheta_0^N,\dots,\hat\vartheta_p^N)^T$ of $\vartheta=(\vartheta_0,\dots,\vartheta_p)^T$ is characterized by the linear system
\begin{align}\label{eq:normalequations}
    A_N\hat\vartheta_N = b_N,
\end{align}
with components given by
\begin{align}
    A_{N,ij} &= \int_0^T\int_\cD F_i(X)_NF_j(X)_N\diff x\diff t, \label{eq:ls_A} \\
    b_{N,i} &= \int_0^T\int_\cD(\partial_tX_N-F_*(X)_N)F_i(X)_N\diff x\diff t, \label{eq:ls_b}
\end{align}
where we write $F_0(X)=\Delta X$ in order to unify notation. This implicitly assumes that the linear model \eqref{eq:Flinearmodel} does not contain redundant basis functions, i.e. the matrix $A_N$ is invertible.
An important special case is the ``plain diffusivity estimator''
\begin{align}\label{eq:plaindiffusivityestimator}
    \hat\vartheta_0^N = \frac{\int_0^T\int_\cD (\Delta X)_N(t, x)\partial_tX_N(t, x)\diff x\diff t}{\int_0^T\int_\cD (\Delta X)_N(t, x)^2\diff x\diff t},
\end{align}
which is obtained by assuming $F=0$ in \eqref{eq:model}. \\

Let us finally discuss a general phenomenon in the derivation of our estimator that is particular to the least squares method for stochastic differential equations. As it turns out, the term $\int_0^T\int_\cD (\partial_tX_N)^2\diff x\diff t$ is not a finite quantity. Consequently, one may argue that the optimization problem stated in \eqref{eq:leastsquares} is ill-posed, because the functional is just an infinite constant. However, in this case there is an easy renormalization procedure: Expanding the square in the functional stated there, we may just drop the ``forbidden'' term since it does not depend on any unknown parameter.\footnote{In the mathematical literature, it is common to derive the estimator from a maximum likelihood approach using Girsanovs theorem (as stated e.g. in \cite[Section 7.6.4]{LiptserShiryayev77}). In this case, the renormalization is implicit.} 
(Note that all other terms can be given a rigorous meaning in the sense of It\^o or Lebesgue integration.) This procedure does not alter the resulting normal equations \eqref{eq:normalequations} or the estimator, which is well-defined even if the functional in \eqref{eq:leastsquares} is not. For that reason, we still call $\hat\vartheta_N$ a least squares estimator. As long as \eqref{eq:normalequations} is used to calculate the estimator this does not make a difference, but for the extension of the method described in Section \ref{sec:ModelSelection}, it is implicitly understood that the least squares functional is renormalized as described.

\subsection{Correlated Noise}

If the noise $\xi$ is correlated, a weighted least squares method should be used instead. For example, in the spatial correlation model \eqref{eq:kernelFracLaplacian}, this means that a power of the Laplacian operator $(-\Delta)^{\gamma}$ should be applied to all terms within the bracket of \eqref{eq:leastsquares}.
In this case, the plain diffusivity estimator \eqref{eq:plaindiffusivityestimator} will be substituted by
\begin{align}\label{eq:correlatednoise}
    \hat\vartheta_0^N = -\frac{\int_0^T\int_{\cD}((-\Delta)^{1+\gamma}X)_N(t, x)((-\Delta)^{\gamma}\partial_tX)_N(t, x)\diff x\diff t}{\int_0^T\int_{\cD}((-\Delta)^{1+\gamma}X)_N(t, x)^2\diff x\diff t}
\end{align}
in that setting.
Note that for $\gamma\notin\N$, the action of $(-\Delta)^\gamma$ is defined as the convolution with the kernel $b_\gamma(x, y)$, which is given as in \eqref{eq:kernelFracLaplacian}.

\subsection{A Possible Extension for Model Selection}\label{sec:ModelSelection}

While least squares type estimators are natural and enjoy good theoretical properties, there are many situations where it may not be a good idea to use them in plain form. As an example, we discuss the problem of model selection with a high-dimensional parameter space.

Consider the situation that we observe some phenomenon in the data (such as bistability or traveling waves), but there are different ways to write down a reaction model that generates such patterns (using e.g. different polynomials, piecewise linear functions or linear combinations of other basis functions such as trigonometric polynomials). 
If we consider a dictionary of various such possible functions for our reaction term $F$, we would expect the vector of coefficients to be \textit{sparse}, i.e. only few entries should be different from zero. Therefore, apart from only estimating the parameters, it is of independent interest to identify the subset of non-zero coefficients in order to decide for the best reaction term.

In the classical statistical learning setup, such a problem can be tackled with the LASSO \cite{Tibshirani96}, which is implicitly defined by adding a regularizing term of the form $\lambda\sum_{i=1}^p|\vartheta_i|$ to the quadratic functional in \eqref{eq:leastsquares}. Here, $\lambda>0$ is a hyperparameter that governs the model complexity. A general discussion can be found in \cite{ElementsOfStatisticalLearning09}.
Other approaches are possible. For example, in \cite{Sindy2016}, an iteratively thresholded least squares estimator is used to account for sparsity in a related setting.

\begin{backgroundbox}
    As in classical statistical learning, the least squares estimator may serve as a baseline for more complex algorithms such as the LASSO, which can be used for model selection.
\end{backgroundbox}

However, up to now, such methods have not been analyzed from a mathematical perspective in the context of stochastic partial differential equations.

\section{Properties of the Statistical Model}\label{sec:StatProp}

In the previous section, we derived estimators for the unknown model parameters from the realization of an approximate process $X_N(t, x)$. In a next step, we will discuss how the 
precision of the estimators improves as $N$ grows, i.e. as the resolution gets finer. Here we work mainly within the above mentioned \textit{spectral approach}.
There is much literature on this approach (see \cite{Lototsky09} and references therein), given that it is conceptually and technically rather simple compared to other approaches that are available, but the main aspects transfer also to the case of local \cite{AltmeyerReiss21, AltmeyerTiepnerWahl22} and discrete \cite{HildebrandtTrabs21} observations.
Recall that in this setting, we assume to observe only $N$ spatial eigenmodes of $X$ as in \eqref{eq:spectralapproach} sampled continuously in time, i.e. there is no discretization error in time. This is a reasonable approximation for high-frequency data (or long time series, if the characteristic time scale for the phenomena described by the SPDE is large). 
Also, unless otherwise stated, we assume a noise model as in \eqref{eq:kernelFracLaplacian}. \\

The mathematical theory of statistics for SPDEs has become a vivid field in the last years (see e.g. the survey \cite{Cialenco18} and references therein).
By now, there is some literature on quantitative evaluation of statistical methods for nonlinear SPDEs, although this field is far from being saturated. We highlight \cite{Huebner93} for an early attempt to find a general framework, \cite{DuncanMaslowskiPasikDuncan00, GoldysMaslowski02} for a study of the ergodic case, \cite{CialencoGlattHoltz11} on the Navier--Stokes equation as well as the more recent works \cite{HildebrandtTrabs21_nonparametric, GaudlitzReiss22}. Stability properties of diffusivity estimation under misspecification as exposed in Section \ref{sec:robust:reaction} below have been systematically studied in \cite{PasemannStannat20, AltmeyerCialencoPasemann23, PasemannFlemmingAlonsoBetaStannat21, Pasemann21}.

\subsection{Estimation Error vs. Spatial Resolution}\label{sec:analysis:N}

As it turns out, reaction and diffusion parameter behave very differently as $N$ grows, and this behaviour is even dimension-dependent. To illustrate this point, let us consider a model problem with a linear source term ($p=1$ and $F_1(X)=X$ in \eqref{eq:Flinearmodel}, i.e. $F_\vartheta(X)=\vartheta_1X$), such that the resulting SPDE is, in fact, linear. This model has been studied in \cite{HubnerKhasminskiiRozovskii93, HuebnerRozovskii95, Lototsky03}. The diffusivity estimator is consistent in the sense that
\begin{align}\label{eq:rate:diffusivity}
    \hat\vartheta_0^N = \vartheta_0 + \mathcal{O}_\mathbb{P}(N^{-\frac{1}{2}-\frac{1}{d}})
\end{align}
as $N$ gets large, i.e. $N^{\frac{1}{2}+\frac{1}{d}}(\hat\vartheta_0^N-\vartheta_0)$ is bounded in probability.\footnote{In the mathematical literature, this is also called \textit{tightness} and defined as $\lim_{M\rightarrow\infty}\sup_{N}\mathbb{P}(|N^{\frac{1}{2}+\frac{1}{d}}(\hat\vartheta_0^N-\vartheta_0)|>M)=0$.}
In particular, the diffusivity can be identified in a finite time horizon $T$ from increasing the spatial resolution.
In contrast, the reaction rate $\vartheta_1$ can be recovered as
\begin{align}\label{eq:rate:reaction}
    \hat\vartheta_1^N = \vartheta_1 + \mathcal{O}_\mathbb{P}(N^{-\frac{1}{2}+\frac{1}{d}})
\end{align}
as $N$ gets large, provided that $d\geq 3$. The estimator still converges to $\vartheta_1$, 
but at a much slower rate.
In dimension $d=2$, the reaction rate can still be consistently estimated, but only with logarithmic convergence rate: $\hat\vartheta_1^N=\vartheta_1+\mathcal{O}_\mathbb{P}(\ln(N)^{-1})$. However, in the one-dimensional setting $d=1$, the estimation error of $\hat\vartheta_1^N$ need not even shrink as $N$ grows! \\

What do the convergence rates \eqref{eq:rate:diffusivity}, \eqref{eq:rate:reaction}, which are stated in terms of ``spectral resolution'' $N$, tell us about a data frame build of ($d$-dimensional) pixels? Assume that the side of each pixel has length $h>0$, such that the size (in pixels) of a frame is of order $h^{-d}$. Using a discrete Fourier transform, the number of eigenmodes obtained from the data and the number of pixels should be of the same order: $N\sim h^{-d}$. This suggests that according to \eqref{eq:rate:diffusivity}, diffusivity can be estimated at rate $h^{1+d/2}$: For example, in $d=2$, if the resolution is increased such that $h$ is cut in half, then the uncertainty of $\hat\vartheta_0^N$ will decrease by a factor of four. In $d=3$, the decay is even faster:\footnote{But of course, cutting $h$ in half in dimension three requires more pixels than in dimension two.}
By making $h$ four times as small we would obtain a decay in uncertainty by a factor of $32$.
In contrast, by \eqref{eq:rate:reaction}, the reaction rate can be estimated only at rate $h^{d/2-1}$: In $d=3$, if the pixel width is made four times as small, then the uncertainty of $\hat\vartheta_1^N$ will decay only by a factor of two!
In $d=2$, the convergence rate is only logarithmic, which will be hard to see in practice.

\subsection{Estimation Error vs. Time Horizon}

Apart from increasing the resolution within each frame, estimation can be made more precise by observing the trajectory for a longer period of time, i.e. by letting the observation time $T$ be large. It is well-known that drift parameters of stochastic differential equations can be estimated at rate $\mathcal{O}_\mathbb{P}(1/\sqrt{T})$, see e.g. \cite{Kutoyants04} for a general discussion. This means that precision is doubled if the observation horizon is extended by a factor of four.
There are various works which aim at similar results for stochastic partial differential equations (see e.g. \cite{KoskiLoges85, Mohapl94, MaslowskiTudor13, KrizMaslowski19}). 
It is important to note that there is no separation between diffusion and reaction terms in the long-time asymptotic regime, i.e. their errors are of the same order if $N$ is fixed and $T\rightarrow\infty$. This is a fundamental contrast to the setting of high spatial resolution ($N\rightarrow\infty$). \\

A precise understanding of the convergence rates of the estimators in different asymptotic regimes helps to answer the following question: 
If I can increase my budget, should I observe more frames in an experiment or rather increase the resolution within each frame? In the present framework, this turns out to depend heavily on 
the parameter to be estimated (reaction vs. diffusion).

\begin{backgroundbox}
    \textbf{Strategies to reduce the estimation error of different terms:} \newline
    
    \begin{tabular}{l|cc}
        & \;\;increase spatial resolution\;\; & \;\;extend time horizon\;\; \\
        \hline
        \textit{diffusivity} & $\checkmark$ & $\checkmark$ \\
        \textit{reaction terms}\;\; & ($\times$) & $\checkmark$
    \end{tabular}
    \vspace{0.3cm}
    
    Higher spatial resolution helps to identify reaction terms in $d\geq 2$, but the gain is small (only logarithmic in dimension two).
\end{backgroundbox}

\subsection{Uncertainty Quantification}\label{sec:UncertaintyQuantification}

In certain situations, we have a central limit theorem (CLT). In this discussion, we restrict to the diffusivity estimator $\hat\vartheta^N_0$, which should be considered a fundamental quantity due to its stability properties (explained below in Section \ref{sec:robust:reaction} and \ref{sec:robust:noise}).
Under conditions that we explain below, we have
\begin{align}\label{eq:CLT}
    N^{\frac{1}{2}+\frac{1}{d}}\left(\hat\vartheta_0^N-\vartheta_0\right)\xrightarrow{d}\mathcal{N}\left(0, \frac{2\vartheta(d+2)}{T\Lambda d}\right)
\end{align}
as $N$ gets large, where ``$\xrightarrow{d}$'' denotes convergence in distribution, and $\Lambda=\lim_{k\rightarrow\infty}\lambda_k/k^{2/d}$ is the proportionality constant of the Laplacian eigenvalues on $\mathcal{D}$. This constant can be identified \cite{Weyl11, Shubin01} as $\Lambda=4\pi^2(B_d|\mathcal{D}|)^{-2/d}$, where $B_d$ is the volume of the $d$-dimensional unit ball, and $|\mathcal{D}|$ is the volume of $\mathcal{D}$ (i.e. the area in $d=2$).
From \eqref{eq:CLT}, it follows that for large $N$, the diffusivity estimator is approximately normally distributed with mean at the true parameter:
\begin{align}
    \hat\vartheta_0^N \approx \vartheta_0 + \sqrt{\frac{2\vartheta(d+2)}{T\Lambda d \,N^{1 + 2/d}}} \mathcal{N}\left(0, 1\right),
\end{align}
and this approximation can be used in a standard way to construct confidence intervals. For example, the interval
\begin{align}
\label{ConfidenceInterval}
    C_N = \left[\hat\vartheta_0^N - 1.96\sqrt\frac{2\vartheta(d+2)}{T\Lambda d\,N^{1+2/d}},\; \hat\vartheta_0^N + 1.96\sqrt\frac{2\vartheta(d+2)}{T\Lambda d\,N^{1+2/d}}\right]
\end{align}
is (for large $N$) an approximate $95\%$ confidence interval. \\

A simple sufficient condition for the CLT \eqref{eq:CLT} to hold is the absense of a reaction term, i.e. $F_\vartheta(X)=0$. In this case, $X(t, x)$ is just the solution to a stochastic heat equation. A more interesting condition depends on the \etaName{} of $F_\vartheta(X)$, which we introduce in the next section.

\subsection{Misspecification of Reaction Terms}
\label{sec:robust:reaction}

We have seen that reaction terms are intrinsically harder to estimate than 
diffusivity: Increasing the spatial resolution will give a comparatively 
small benefit in terms of uncertainty reduction, if at all. This 
observation, however, has a positive side effect: 
It goes hand in hand with the fact that the diffusivity estimate 
is rather \textit{stable} under wrongly specified reaction models. \\

We can quantify the stability as follows: 
In the model \eqref{eq:model}, we associate to $F_\vartheta(X)$ a number 
$\eta>0$, called the \textit{\etaName}\footnote{For a precise mathematical discussion, see \cite[Section 2.2]{Pasemann21}.} of $F$. Roughly speaking, $\eta$ is 
two minus the order of $F$ as a differential operator. So, a reaction term 
$f(X)$ has \etaName{} $\eta=2$, whereas an advection term $\nabla\cdot(Xv)$ 
has the \etaName{} $\eta=1$. If $F_\vartheta(X)$ has several components as 
in \eqref{eq:Flinearmodel}, its \etaName{} $\eta=\min\{\eta_1, \dots, 
\eta_p \}$ is just the smallest stability of all its components. \\

Now assume that the data $X$ is generated from an SPDE that contains the true reaction term $F$ (with \etaName{} $\eta>0$), whereas $\hat\vartheta_0^N$ is constructed from a wrongly specified dynamical model that includes instead the term $\tilde F$ (with the same or larger \etaName{}).
For example, a local source term may be replaced by another such term (possibly with a different bifurcation structure), or within an advection term, the velocity field may be exchanged. This includes the case where $F$ is just dropped, i.e. formally substituted by zero. Now the question is: What can we say about inference on diffusivity if $X$ comes from the true model, but the estimator $\hat\vartheta_0^N$ is built from its misspecified version?
The following Theorem \ref{thm} now in particular states lower bounds 
on the stability index, up to which the 
convergence of 
$\hat{\vartheta}_0^N$, 
as well as associated confidence intervals, are not affected.

\begin{theorem}[\cite{Pasemann21}] \
\label{thm}
\begin{enumerate}
    \item If
\begin{align}\label{eq:exregdimension}
    \eta \geq 1 + \frac{d}{2},
\end{align}
then the convergence of $\hat\vartheta_0^N$ as stated in \eqref{eq:rate:diffusivity} remains unchanged.\footnote{Technically, if \eqref{eq:exregdimension} holds with equality, the guaranteed convergence rate loses an arbitrarily small polynomial factor, e.g. for $\eta=d=2$ we have $\hat\vartheta_0^N=\vartheta_0 + \mathcal{O}_\mathbb{P}(N^{-1+\varepsilon})$ for any $\varepsilon>0$. 
}
    \item If \eqref{eq:exregdimension} holds strictly, i.e. $\eta>1+d/2$, then $\hat\vartheta^N_0$ even satisfies the CLT \eqref{eq:CLT}, leading to identical confidence intervals \eqref{ConfidenceInterval}.
    \item Otherwise, if $0<\eta<1+d/2$, then \eqref{eq:rate:diffusivity} is replaced by\footnote{Again, here we neglect an arbitrarily small polynomial factor in the rate.}
\begin{align}\label{eq:RateExcessRegularity}
    \hat\vartheta_0^N = \vartheta_0 + \mathcal{O}_\mathbb{P}(N^{-\frac{\eta}{d}}).
\end{align}
\end{enumerate}
\end{theorem}

This has two implications: First, the estimator 
still converges as $N$ grows. This is a universal property that holds true for any $F$ with positive \etaName. 
Second, if $\eta$ is large enough as in \eqref{eq:exregdimension}, then not even the convergence rate is affected, i.e. 
we can still expect the same gain in precision as $N\rightarrow\infty$, even if the underlying reaction model is wrong. However, for small $\eta$, the misspecified model lowers the rate of convergence. In this case, a larger resolution is needed in order to obtain the same precision as in the correctly specified case.

Interestingly, the stability of $\hat\vartheta^N_0$ depends on the dimension $d$, with better rates in low dimensions. For example, a wrong reaction term with $\eta=2$ will not affect the convergence speed of $\hat\vartheta_0^N$ in dimensions one and two. In $d=1$, we will even keep the central limit theorem for $\hat\vartheta_0^N$ with identical confidence intervals \eqref{ConfidenceInterval}, which is explained in Section \ref{sec:UncertaintyQuantification}.

An immediate consequence of the previous discussion is that a wrong advection model affects the diffusivity inference more severely than a wrong reaction model.

\begin{backgroundbox}
    In the high resolution limit $N\rightarrow\infty$, the diffusivity estimator $\hat\vartheta^N_0$ is stable under misspecification in the reaction model.
\end{backgroundbox}

Note that a similar stability cannot be expected in the large time regime 
where all estimators will converge at the same rate $\mathcal{O}_\mathbb{P}
(1/\sqrt{T})$. Thus, increasing the observation horizon will not weaken the 
impact of model misspecification. 

We close this section with a list of commonly used terms and their stability index:\footnote{All examples are discussed in \cite[Section 2.4]{Pasemann21} except for the inhibitor term \eqref{eq:inhRep}, whose stability index is implicit in the proof of \cite[Proposition 1]{PasemannFlemmingAlonsoBetaStannat21}.}

\begin{center}
\begin{tabular}{lc|l}
    \multicolumn{2}{c|}{\textbf{model for $F(X)$}} & \;\;\textbf{\etaName}\;\; \\
    \hline
    \;\;\textit{reaction term} & $f(X)$ \;\;& \quad\quad$\eta=2$ \\
    \;\;\textit{advection term} & $\nabla\cdot(Xv)$ \;\;& \quad\quad$\eta=1$ \\
    \;\;\textit{inhibitor term} as in \eqref{eq:inhRep} &  & \quad\quad$\eta=4$ \\
    \;\;\textit{fractional diffusion} & $(-\Delta)^\frac{\alpha}{2} X$ \;\;& \quad\quad$\eta=2-\alpha$
\end{tabular}
\end{center}

In align with the discussion in Section \ref{sec:analysis:N}, a large stability index should be seen as an indicator that the corresponding term is harder to estimate.

\subsection{Misspecification of Noise Terms}\label{sec:robust:noise}

Apart from the reaction term, let us study the robustness of results on $\hat\vartheta_0^N$ concerning misspecification of the noise model for $\xi$. The theory outlined above has been derived under the assumption that $\xi$ is white in time, and possibly colored in space, where the correlation kernel $b_{-\gamma}$ is given by \eqref{eq:kernelFracLaplacian}. This noise model contains two parameters, namely the noise intensity $\sigma$ as well as the degree of spatial correlation $\gamma$.
\\

\textbf{The case of wrongly specified parameters.} In order to construct $\hat\vartheta^N_0$, no knowledge on $\sigma$ is required. On the other hand, $\gamma$ does appear in \eqref{eq:correlatednoise}. However, if the true exponent $\gamma$ is replaced by some $\alpha\neq\gamma$ when deriving $\hat\vartheta_0^N$, this does not affect any of the results on convergence rates stated above, as long as $\alpha$ is not too small. (For example, one can always choose $\alpha\geq\gamma$.)

We mention that at least in theory, $\sigma$ and $\gamma$ can be identified from continuous time observations 
of $X$ via its quadratic variation. More precisely, for the first eigenmode $X_N=X_1$ of $X$, given by \eqref{eq:spectralapproach}, 
we have
\begin{align}
    \lim_{k\rightarrow\infty}\sum_{i=1}^k\int_\cD\left(X_1\left({\frac{Ti}{k}}, x\right) - X_1\left({\frac{T(i-1)}{k}}, x\right)\right)^2\diff x = \sigma^2\lambda_1^{-2\gamma}T,
\end{align}
where $\lambda_1$ is the first eigenfrequency of the domain $\cD$, and the limit refers to convergence in probability. A similar equation holds for the second eigenmode of $X$ (with the second eigenfrequency $\lambda_2$ appearing in the limit), and once these two limits are known, the unknown noise parameters $\sigma$ and $\gamma$ can be easily recovered.
\\

\textbf{The case of a wrong parametric model.} Next, consider Ornstein--Uhlenbeck noise as in Example \ref{ex:noise:OUtwo}. Here we induce some slight temporal correlation to the true data generating model. Neglecting this correlation when deriving the estimator turns out to be as severe as a reaction model misspecification ($\eta=2$). This means, according to \eqref{eq:exregdimension} and the discussion thereafter, that the convergence rate of $\hat\vartheta_0^N$ is not affected in $d=1$ or $d=2$, and for $d\geq 3$ it shrinks to the term given by \eqref{eq:RateExcessRegularity}. A detailed discussion of this setting can be found in \cite[Chapter 3]{Pasemann21}.

\subsection{A Counterexample}

We have seen that the diffusivity estimator $\hat\vartheta^N_0$ is robust against misspecification in the drift, as well as certain deviations from the white noise model. However, this cannot (and does not) apply to all possible kinds of misspecification. A particularly revealing example is the noise model from Example \ref{ex:noise:OUone}, where $\xi$ is modelled as an Ornstein--Uhlenbeck process. 
We highlight this example, because it is a very natural one: Bearing in mind that the movement of a classical Brownian particle can be described by the Wiener process or an integrated Ornstein--Uhlenbeck process (where the velocity of the particle is modelled as an Ornstein--Uhlenbeck process), the analogous model for $\xi$ would be either space-time white noise, or the model from Example \ref{ex:noise:OUone}. However, if the latter is the true data-generating model for $X$, this would break 
the stability of $\hat\vartheta^N_0$ - in fact, this estimator would not even converge to the true parameter: 
One can prove that in this setting, even in the simplest case $\mu=0$ the estimator $\hat\vartheta_0^N$ converges to zero as $N$ grows! So what went wrong? The answer is related to the smoothness of $X$. Note that an integrated Ornstein-Uhlenbeck process has a well-defined derivative, while a trajectory of a Wiener process has not. This observation transfers to our setting: A field $X$ having a well-defined derivative $\partial_tX$ (in the classical sense) is ``too regular'' to be well approximated by a white noise model.\footnote{Note that in this particular example, the problem can be mitigated by replacing $X$ by $Y=\partial_tX$ as the fundamental quantity, i.e. writing down the induced dynamical model for $Y$ and using a least squares method based on that model. See \cite[Chapter 3.2]{Pasemann21} for details.}

\section{Notes on Implementation}\label{sec:Implementation}

So far, we focused on the situation that the space-time random field $X$ is approximated by some spatially discretized process $X_N$. In most situations, however, it is reasonable to assume that the data is given as a discrete approximation not only in space, but also in time. In this case, it is straightforward to discretize the integrals appearing in the definition \eqref{eq:normalequations} of the least squares estimator $\hat\vartheta_N$. There is, however, a caveat for the terms that include the time derivative $\partial_tX_N$, which is related to the fact that mathematically, the field $X_N$ is not differentiable in time. More precisely, consider an integral of the form
\begin{align}\label{eq:stochint}
    \int_0^T\int_\cD F(X)_N\partial_tX_N\diff x\diff t.
\end{align}
Terms of that form arise in \eqref{eq:ls_b} or the numerator of \eqref{eq:plaindiffusivityestimator}. These integrals have to be understood in the sense of It\^o, which means that they arise as the limit in probability of sums of the form
\begin{align}
    \sum_{i}\int_\cD F(X)_N(t_i, x)\frac{X_N(t_{i+1}, x)-X_N(t_i, x)}{t_{i+1}-t_i}\diff x\;(t_{i+1}-t_i).
\end{align}
Here, it is crucial that the term $F(X)_N(t_i, x)$ is evaluated at the left boundary point $t_i$ of the interval $[t_i, t_{i+1}]$. 
The reason is that otherwise additional correlation between $F(X)_N$ and the difference quotient is introduced, which will shift the final result.

This is particularly relevant when simulating such integrals (in this case, the mid-point rule or related higher order quadrature methods known from the approximation of a classical Riemann integral do not work without further modification) but also plays a role when approximating these integrals from discretely given data, always under the standing assumption that the driving noise $\xi$ is (close to) white in time. \\

In some situations it is possible to avoid the direct approximation of an It\^o integral by transforming it with It\^o's formula (see e.g. \cite{LeGall16}). For example, within the spectral approach \eqref{eq:spectralapproach}, the numerator of the ``plain diffusivity estimator'' \eqref{eq:plaindiffusivityestimator} can be written as follows:
\begin{align}\label{eq:NumeratorItoFormula}
    \int_0^T\int_{\cD}(\Delta X)_N(t, x)\partial_tX_N(t, x)\diff x\diff t = \frac{1}{2}\sum_{k=1}^N\lambda_k\left((X^{(k)}_T)^2-(X^{(k)}_0)^2-T\right),
\end{align}
where $\lambda_k$ is the $k$-th eigenfrequency of the Laplacian $\Delta$ on $\cD$, and $X^{(k)}_t=\int_\cD X(t, y)\Phi_k(y)\diff y$ is the $k$-th mode of $X(t, x)$.
The right-hand side of \eqref{eq:NumeratorItoFormula} allows for explicit calculation. We highlight the It\^o correction term $-T$ appearing therein, which is related to correlation appearing in the 
square of the stochastic process
$X^{(k)}_t$.

\section{Application to \textit{D. discoideum} giant cell data}
\label{sec:Application} 

We finally show how our method can be applied to measurements of actin concentration within \textit{D. discoideum} giant cell data.
The results presented here are taken from \cite{PasemannFlemmingAlonsoBetaStannat21}. 
We refer to Chapter 4 therein
for a detailed discussion. \\

We assume that the data follow an activator-inhibitor dynamic as in Example \ref{ex:actinh}. Parametrizing the reaction term as in \eqref{eq:actinh:parametrization}, we have three reaction parameters $\vartheta_1,\vartheta_2,\vartheta_3$ in addition to the unknown diffusivity $\vartheta_0$. The reaction parameters may be treated as known or unknown, leading to a more rigid or flexible statistical model. \\

We compare four diffusivity estimators, which we apply to the data. 
\begin{itemize}
    \item $\hat\vartheta_0^{\mathrm{lin}, N}$ is just the estimator \eqref{eq:plaindiffusivityestimator}, where $\vartheta_1, \vartheta_2, \vartheta_3$ are formally set to zero. 
    \item $\hat\vartheta_0^{2, N}$ is implicitly defined by \eqref{eq:normalequations}, where $\vartheta_0,\vartheta_1$ are treated as unknown, and the remaining parameters $\vartheta_2,\vartheta_3$ are fixed.
    \item $\hat\vartheta_0^{3, N}$ treats $\vartheta_0,\vartheta_1,\vartheta_2$ as unknown and $\vartheta_3$ as fixed.
    \item $\hat\vartheta_0^{4, N}$ treats all parameters $\vartheta_0,\vartheta_1,\vartheta_2,\vartheta_3$ as unknown.
\end{itemize}

\begin{figure}
    \includegraphics[width=0.49\textwidth]{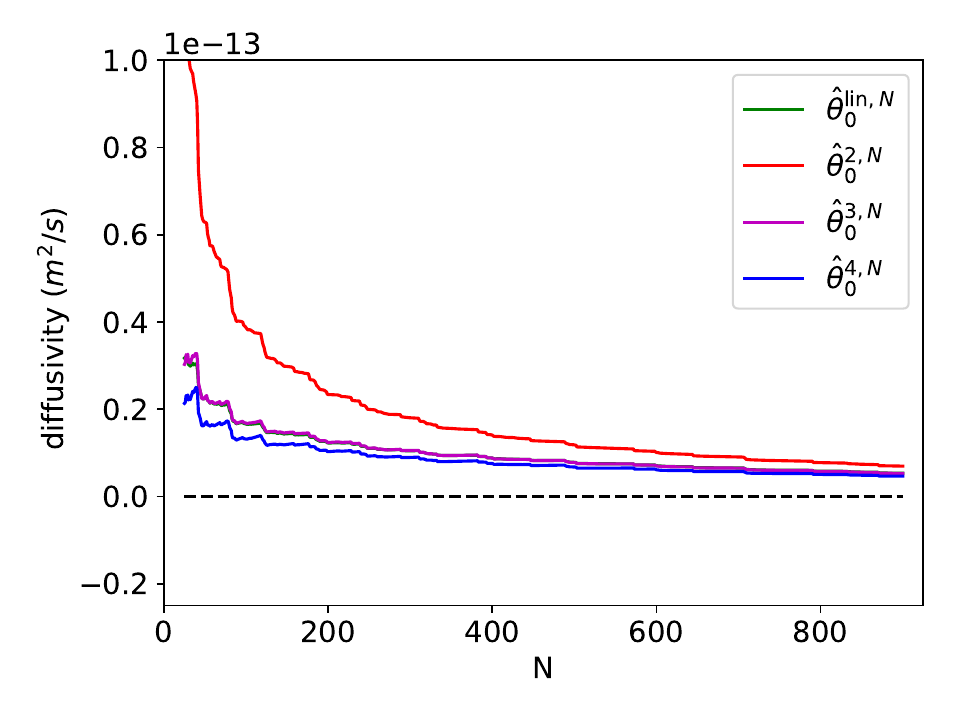}
    \includegraphics[width=0.49\textwidth]{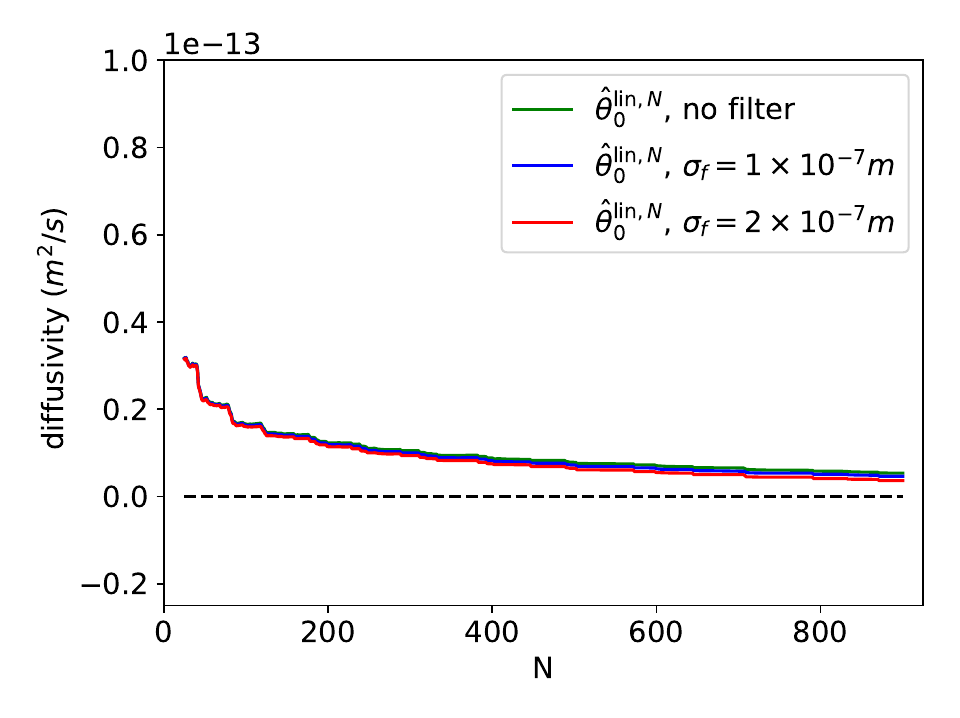}
    \caption{
    Plots of $\hat\vartheta_0^N$ vs. $N$ for different diffusivity estimators $\hat\vartheta_0^N$ constructed from experimental data, for $N\geq 25$. \textit{(Left)} Different assumptions are made on what parameters are considered known or unknown. \textit{(Right)} Invariance of diffusivity estimation under preprocessing the data with kernel smoothing in space.
    }\label{fig:est:diff}
\end{figure}

The left panel of Figure \ref{fig:est:diff} shows the result for a particular measurement of a cell. Interestingly, $\hat\vartheta_0^{\mathrm{lin}, N}$ and $\hat\vartheta_0^{3, N}$ are almost indistinguishable: The detailed reaction model does not seem to contribute to a better understanding of the diffusivity. On the other hand, $\hat\vartheta_0^{2, N}$ deviates from the other estimators for low frequencies (in this sense, fixing all reaction parameters may be ``too rigid''), but approaches the other estimators for high frequencies, in accordance with the results from Section \ref{sec:robust:reaction}. \\

One may ask if the assumption that the data is generated by a stochastic reaction-diffusion model is reasonable at all. A first sanity check can be carried out based on the following observation: If the solution to a heat equation is smoothed (in space) with some kernel $K(x)$, the resulting field still solves a heat equation.\footnote{The reason is that the mathematical operations ``convolution with a kernel $K(x)$'' and ``application of the Laplacian operator $\Delta$'' commute.} Consequently, if $X(t, x)$ is generated by a ``heat equation up to lower order terms'' (which we identify as reaction and noise terms), then $\tilde X(t, x)=\int_\mathcal{D}K(x-y)X(t, y)\diff y$ still comes from such a perturbed heat equation (although the reaction and noise terms will have changed).
The right-hand side of Figure \ref{fig:est:diff} shows the result of applying $\hat\vartheta_0^{\mathrm{lin}, N}$ to the original cell data as well as to a filtered version of the same data, where a kernel with different values for the bandwidth $\sigma_f$ is applied as a form of preprocessing. The estimated diffusivity changes only very slightly, as is expected if the stochastic reaction-diffusion model is reasonable. \\

\begin{figure}
    \includegraphics[width=0.49\textwidth]{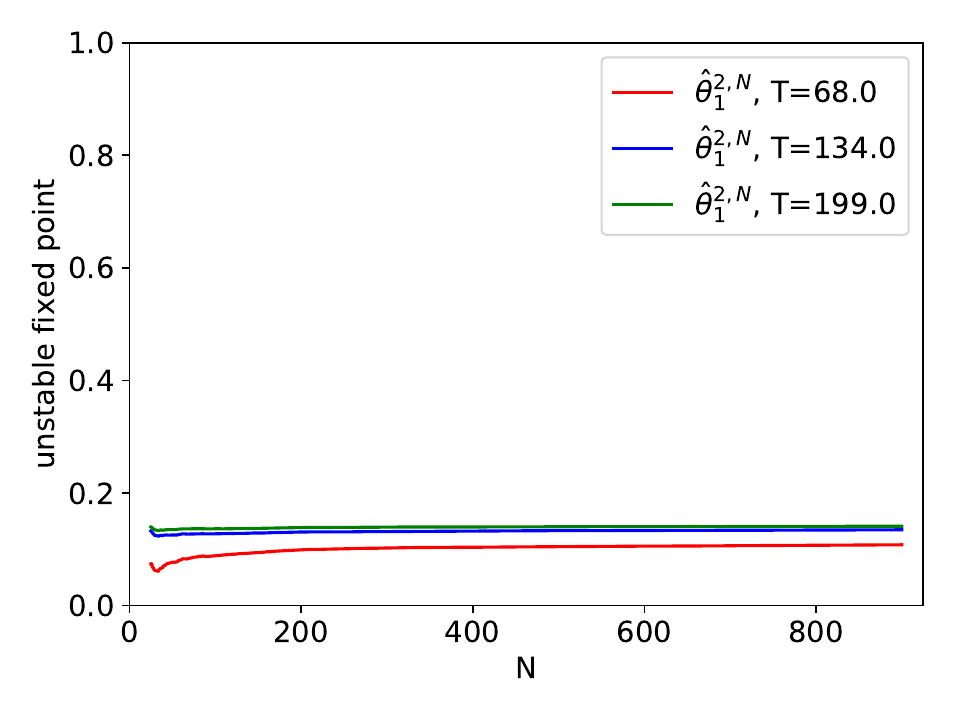}
    \includegraphics[width=0.49\textwidth]{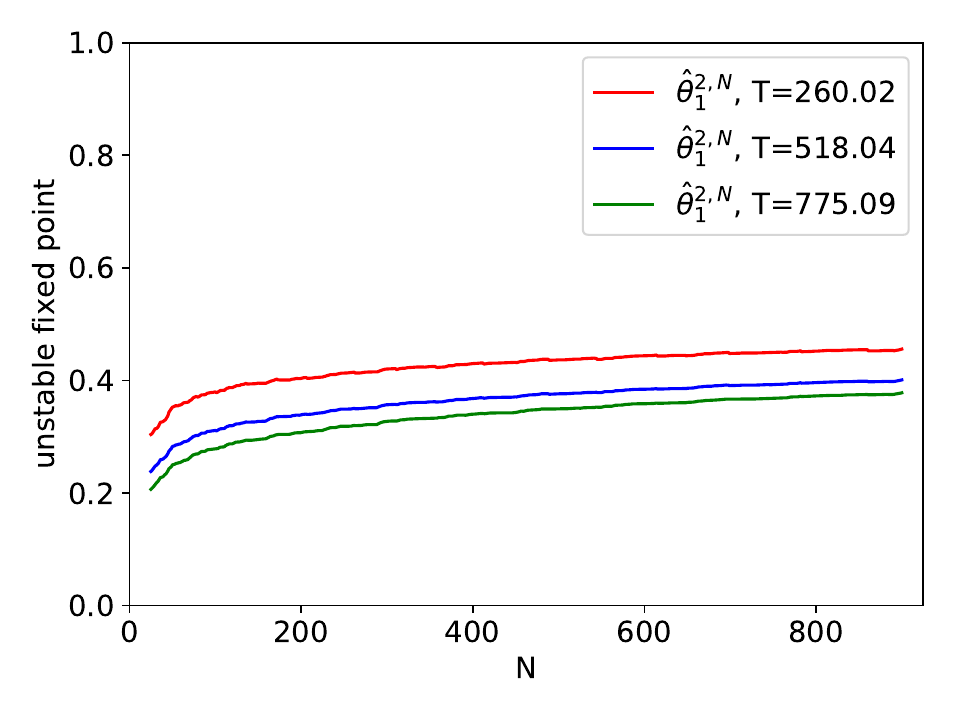}
    \caption{Graphs of $\hat\vartheta_1^{2, N}$ vs. $N$ on \textit{(left)} simulated and \textit{(right)} experimental data, where all frames up to time $T$ are used for estimation. All times are given in seconds. We restrict to $N\geq 25$.
    }\label{fig:est:zero}
\end{figure}

Next, consider the estimator $\hat\vartheta_1^{2, N}$ for $\vartheta_1$, which is constructed from \eqref{eq:normalequations} under the model assumptions that $\vartheta_2, \vartheta_3$ are fixed and known (in analogy to $\hat\vartheta_0^{2, N}$). From \eqref{eq:actinh:parametrization}, $\vartheta_1$ may be identified with the ``effective unstable fixed point'' of the reaction dynamics. Figure \ref{fig:est:zero} shows the result on simulated and experimental data. On the simulations, increasing the number of Fourier modes $N$ has almost no effect on the value of $\hat\vartheta_1^{2, N}$. This is due to the logarithmic convergence rate, as discussed in Section \ref{sec:analysis:N}. On real data, the estimated value does indeed lie between zero and one (these two values correspond to the stable fixed points in the FitzHugh--Nagumo system). Further, it is rather stable in $N$. It may be therefore used as a statistic that reflects an interpretable property of data exhibiting bistability or travelling waves.

\section{Conclusion}

In this work we have developed statistical inference 
methods for parameter estimation in stochastic reaction-diffusion systems 
with a view towards quantitative model evaluation for space-time data 
$X_N (t,x)$. 
We discussed the fundamentally different statistical properties of diffusion and reaction estimation, with special emphasis on the issue of drift misspecification.
As a proof of concept, we have applied our analysis to the 
statistical inference on intracellular actin concentration in the social 
amoeba \textit{D. discoideum}. 
Both the effective diffusivity and the effective unstable fixed point 
appear to be reasonable statistics that reflect inherent properties of the data. They may be used in further research in order to understand and classify spatiotemporal data according to their dynamics.
So far we did not take measurement noise into 
account. In the above case of fluorescence microscopy, e.g., this will add 
independent shot noise and detector noise to $X_N (t,x)$, which 
requires additional filtering. This will be the content of future work.

\section*{Acknowledgements}

This research has been partially funded by the Deutsche Forschungsgemeinschaft (DFG)- Project-ID 318763901 - SFB1294.
We like to thank our coauthors from the work \cite{PasemannFlemmingAlonsoBetaStannat21} Sven Flemming for providing experimental data and Sergio Alonso for sharing numerical code.

\bibliographystyle{amsalpha}

\bibliography{lit,lit_CB}

\end{document}